\documentclass[%
 aip,
 amsmath,amssymb,
 reprint,%
]{revtex4-1}

\usepackage{graphicx}
\usepackage{dcolumn}
\usepackage{bm}

\usepackage[utf8]{inputenc}
\usepackage[T1]{fontenc}
\usepackage{mathptmx}
\usepackage{textgreek}
\usepackage{etoolbox}

\makeatletter
\def\@email#1#2{%
 \endgroup
 \patchcmd{\titleblock@produce}
  {\frontmatter@RRAPformat}
  {\frontmatter@RRAPformat{\produce@RRAP{*#1\href{mailto:#2}{#2}}}\frontmatter@RRAPformat}
  {}{}
}%
\makeatother
\begin{document}

\preprint{AIP/123-QED}

\title[Sample title]{Enhancement-Mode Vertical $\beta$-Ga$_2$O$_3$ U-Trench MOSFET with MOCVD Regrown n$^+$ Contact Layers and Nitrogen-Implanted Current Blocking Layer}
\author{Walid Amir}
\author{Jiawei Liu}
\author{Surajit Chakraborty}
\affiliation{ 
Department of Electrical and Computer Engineering, University at Buffalo, Buffalo, NY 14260 USA
}%

\author{Dongsu Yu}
\author{Md. Mosarof Hossain Sarkar}
\author{Hongping Zhao}
 \affiliation{%
Department of Electrical and Computer Engineering, The Ohio State University, Columbus, OH 43210, US
}%
\author{Uttam Singisetti}
\email{uttamsin@buffalo.edu}
\affiliation{ 
Department of Electrical and Computer Engineering, University at Buffalo, Buffalo, NY 14260 USA
}%
\date{\today}

\begin{abstract}
In this work, an implantation-free ohmic contact technology based on selectively MOCVD-regrown Si-doped n$^+$ layers is demonstrated for enhancement-mode vertical $\beta$-Ga$_2$O$_3$ U-trench MOSFETs. The regrown n$^+$ contact structure eliminates the need for implantation-based ohmic contact formation while maintaining excellent electrical characteristics. A multi-energy nitrogen-ion-implanted current blocking layer (CBL) followed by 1100~$^\circ$C activation annealing in N$_2$ ambient for 30 min was employed to achieve normally-OFF operation. Transmission line model measurements yielded a low specific contact resistivity of $2.65 \times 10^{-7}~\Omega\cdot$cm$^2$. The fabricated devices exhibited a threshold voltage of approximately 5~V, an ON/OFF current ratio of $1.15\times10^{6}$, a peak current density of 158A/cm$^2$, and a specific ON-resistance of 120.9~m$\Omega\cdot$cm$^2$. Three-terminal OFF-state breakdown voltages ranging from 920 to 980~V were achieved at $V_{GS}=0$~V. Multi-finger MOSFETs show current scaling to 0.25 A. These results demonstrate that selectively MOCVD-regrown n$^+$ contact layers provide a promising implantation-free approach for realizing high-performance vertical $\beta$-Ga$_2$O$_3$ power MOSFETs.

\end{abstract}

\maketitle

Ultra-wide-bandgap (UWBG) semiconductors are attracting significant attention for next-generation high-voltage and high-efficiency power electronics due to their ability to achieve high breakdown voltage and low conduction loss. Among them, $\beta$-Ga$_2$O$_3$ has emerged as a promising candidate because of its wide bandgap of approximately 4.8--4.9~eV and high theoretical critical electric field approaching 8~MV/cm, resulting in a projected Baliga’s figure of merit (BFOM) substantially higher than those of SiC and GaN. In addition, large-area $\beta$-Ga$_2$O$_3$ substrates can be fabricated using melt-growth techniques, enabling scalable and relatively low-cost wafer production for future power device applications\cite{Higashiwaki2012,Kuramata2016,Galazka2016,Zhang2019,WongReview2020}.

To fully exploit the high critical electric field of $\beta$-Ga$_2$O$_3$, vertical device architectures are preferred because the breakdown voltage can be increased through drift-layer engineering without sacrificing chip area. Consequently, extensive efforts have been devoted toward developing vertical $\beta$-Ga$_2$O$_3$ devices, including FinFETs, CAVETs, VDBFETs, and U-shaped trench MOSFETs (UMOSFETs). Recent studies have demonstrated remarkable progress in device performance, including multi-fin vertical transistors with breakdown voltages exceeding 10~kV and UMOSFETs with current densities exceeding 700~A/cm$^2$ \cite{Hu2018,WongNandSi2018,WongCAVET2019,Hu2019,Ma2023,Liu2024,Wakimoto2025,Bhattacharyya2021,Sharma2020,Zeng2018,Sharma2022,Luo2025,roy2025multi}. However, realization of high-performance enhancement-mode (E-mode) vertical MOSFETs remains challenging because effective p-type doping in $\beta$-Ga$_2$O$_3$ is currently not feasible.

To overcome this limitation, current blocking layer (CBL)-based architectures have emerged as an effective approach for realizing normally-OFF operation. In these devices, deep acceptor-like states introduced through nitrogen or magnesium doping compensate the background donor concentration and suppress current conduction at zero gate bias \cite{Peelaers2019,Gake2019,Feng2020,SahaAPL2023,ZengMg2022}. Among these approaches, nitrogen-ion-implanted CBLs have attracted considerable attention because of their strong current blocking capability and superior thermal stability \cite{WongNandSi2018,WongCAVET2019,WongAPL2021,Ma2023,Liu2024,HuImplant2025,LiuAPEX2025,Zou2025}. Recent demonstrations of N-ion-implanted UMOSFETs and VDMOSFETs have achieved kilovolt-class breakdown voltages together with enhancement-mode operation through optimized implantation and annealing conditions \cite{WongNandSi2018,WongCAVET2019,Ma2023,Liu2024,HuImplant2025,LiuAPEX2025,Zou2025,Xu2025}, establishing nitrogen-implanted CBLs as a mature and effective solution for normally-OFF vertical $\beta$-Ga$_2$O$_3$ MOSFETs.

Despite these advances, the formation of low-resistance ohmic contacts remains an important challenge for vertical $\beta$-Ga$_2$O$_3$ MOSFETs \cite{Saha2025}. Most reported vertical devices employ high-dose Si-ion implantation followed by high-temperature activation annealing to realize ohmic contacts, increasing process complexity and introducing implantation-induced crystal damage\cite{WongNandSi2018,WongCAVET2019,Ma2023,Liu2024,Xu2025}. An implantation-free contact technology capable of achieving low-resistance ohmic contacts while simplifying fabrication would therefore be highly desirable. Selectively MOCVD-regrown Si-doped n$^+$ contact layers provide a promising alternative by eliminating contact implantation and activation annealing while preserving excellent ohmic characteristics \cite{Meng2022}.

\begin{figure}[t]
\begin{center}
    \includegraphics[width=\linewidth]{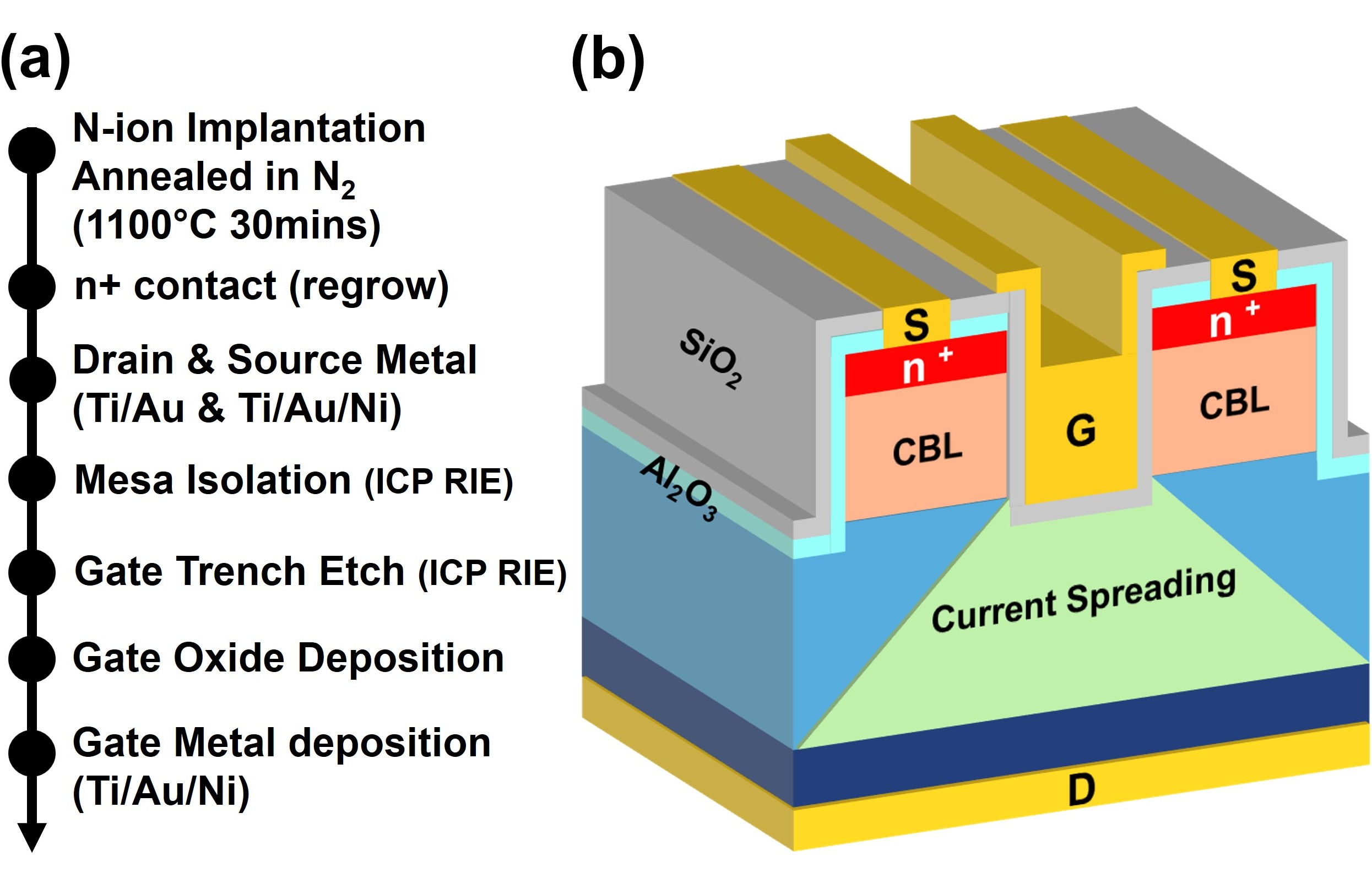}
    \caption{(a) Fabrication process flow of the vertical $\beta$-Ga$_2$O$_3$ U-trench MOSFET. (b) Three-dimensional schematic of the fabricated enhancement-mode vertical $\beta$-Ga$_2$O$_3$ U-trench MOSFET.}
    \label{fig1}
\end{center}
\end{figure}

In this work, we demonstrate an implantation-free source/drain contact technology for vertical $\beta$-Ga$_2$O$_3$ U-trench UMOSFETs using MOCVD-regrown Si-doped n$^+$ layers. The regrown-contact process eliminates contact implantation and high-temperature activation annealing while maintaining excellent ohmic characteristics and enabling high-performance enhancement-mode operation. The fabricated devices exhibited a threshold voltage of approximately 5~V, an ON/OFF ratio of $1.15\times10^6$, a peak current density of 158A/cm$^2$, a specific ON-resistance of 120.9~m$\Omega\cdot$cm$^2$, and OFF-state breakdown voltages approaching 1~kV. These results demonstrate the strong potential of combining nitrogen-ion-implanted CBLs with regrown ohmic contacts for future high-performance $\beta$-Ga$_2$O$_3$ power electronics.

The process flow of the fabricated enhancement-mode vertical $\beta$-Ga$_2$O$_3$ U-trench MOSFET is illustrated in Fig.1. The devices were fabricated on commercially available Sn-doped (001) $\beta$-Ga$_2$O$_3$ substrates with an approximately 10~$\mu$m unintentionally doped drift layer grown by halide vapor phase epitaxy (HVPE). A nitrogen-ion-implanted current blocking layer (CBL) was formed near the surface of the drift region using multi-energy N implantation\cite{WongNandSi2018,WongCAVET2019,Ma2023,Liu2024,HuImplant2025}. Implantation energies of 180, 260, 360, 480, and 600~keV were employed to obtain an approximately box-shaped nitrogen profile extending to a depth of $\sim$900nm, as shown in Fig. 2. The total implantation dose was $2\times10^{14}$~cm$^{-2}$, resulting in a nearly uniform nitrogen concentration in the low-$10^{18}$~cm$^{-3}$ range across the implanted region. The multi-energy implantation approach was adopted to minimize abrupt doping discontinuities and achieve a smoother electric-field distribution under reverse bias conditions.

Following implantation, the samples were annealed at 1100~$^\circ$C in N$_2$ ambient for 30 min to activate the implanted nitrogen-related acceptor states and recover implantation-induced crystal damage \cite{HuImplant2025,LiuAPEX2025,Zhou2022,Zou2025}. The high-temperature annealing process is essential for restoring crystal quality and suppressing defect-assisted leakage current induced during ion implantation. The implanted nitrogen compensates the background donor concentration near the surface of the drift layer, thereby forming an effective current blocking layer that suppresses vertical current conduction at zero gate bias and enables enhancement-mode operation.
\begin{figure}
\begin{center}
    \includegraphics[width=0.8\linewidth]{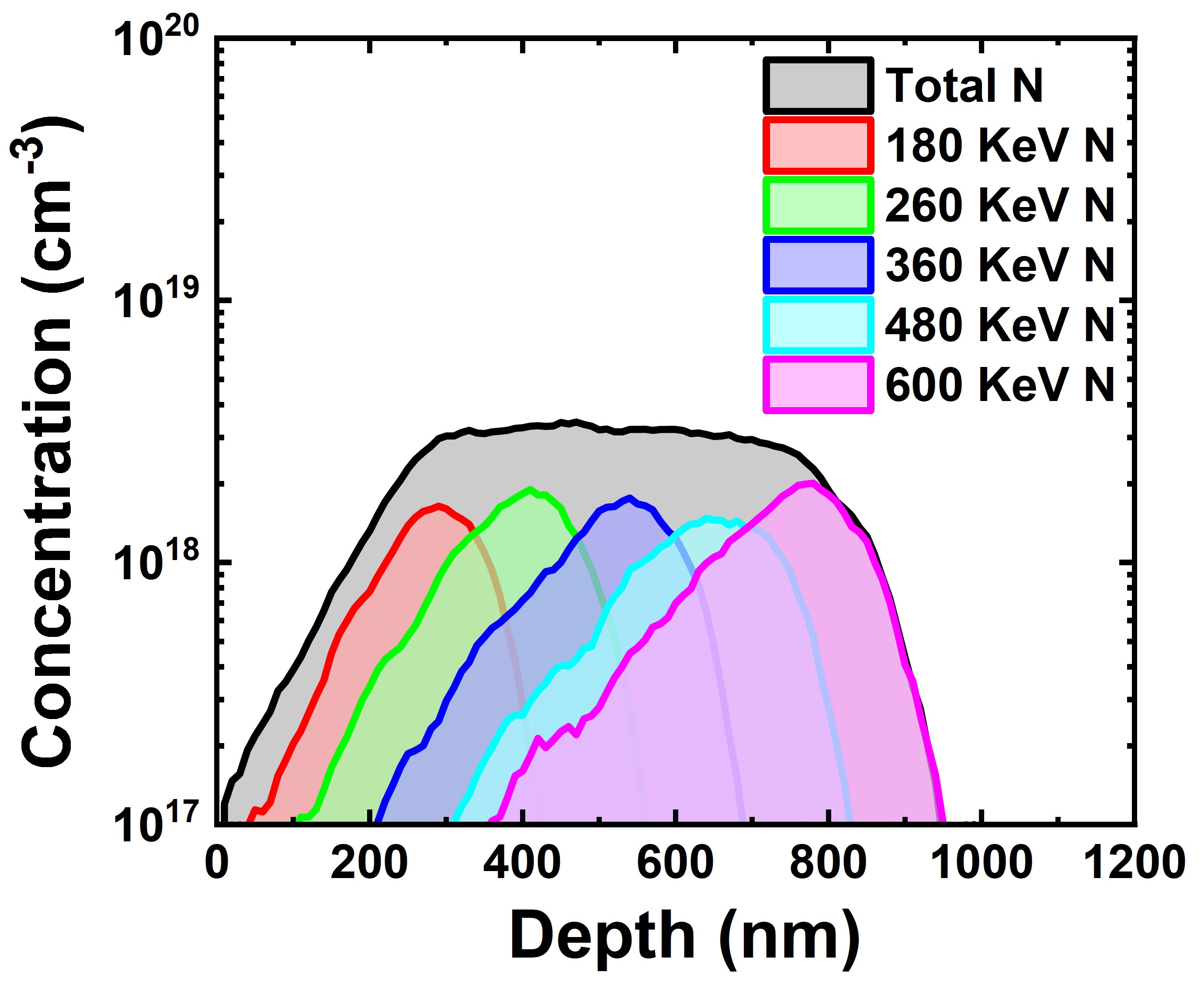}
    \caption{Simulated multi-energy N-ion implantation profiles and total nitrogen concentration forming the current blocking layer (CBL).}
    \label{fig2}
\end{center}
\end{figure}

\begin{figure}[b]
\begin{center}
    \includegraphics[width=\linewidth]{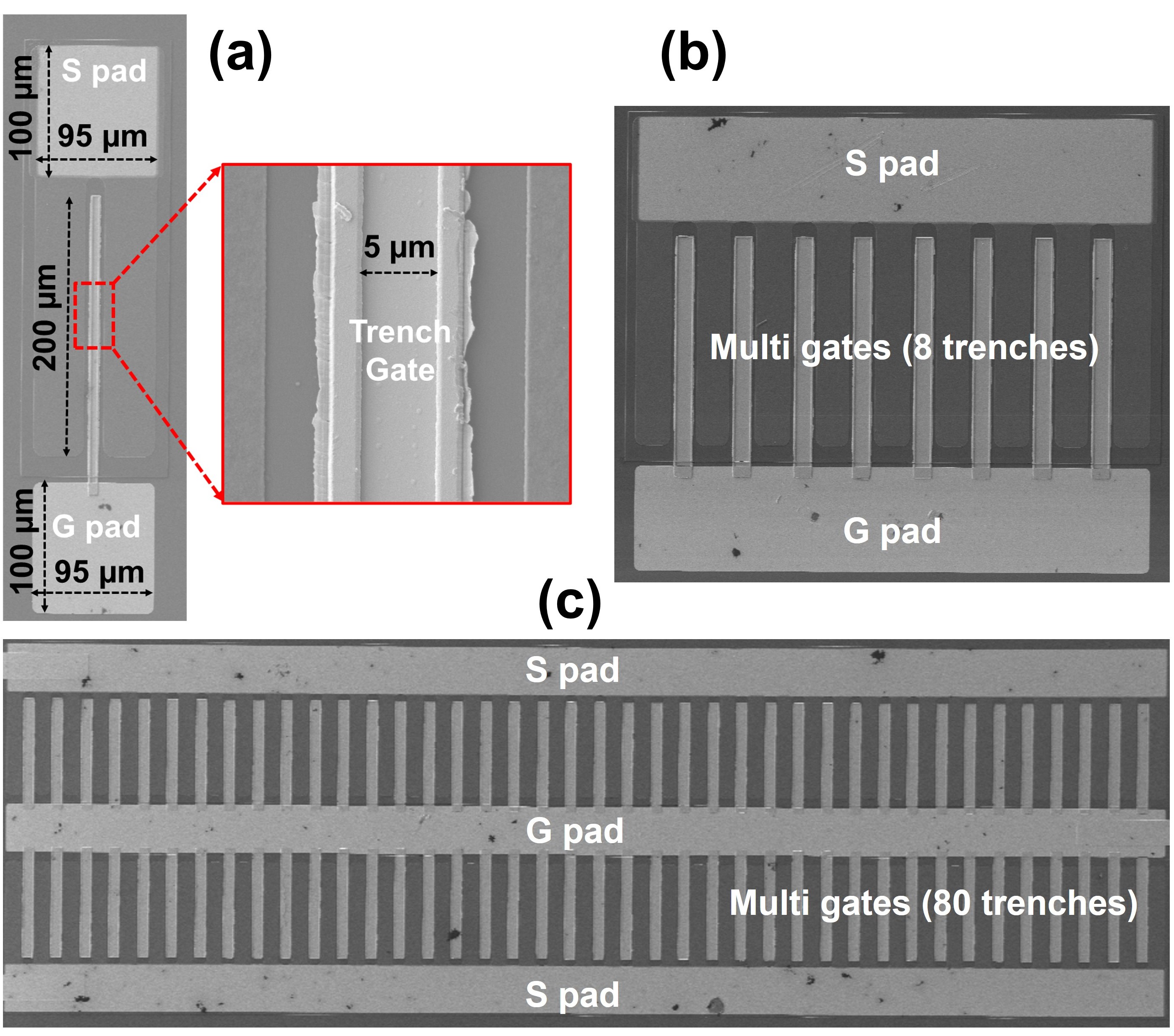}
    \caption{(a) Top-view SEM image of a single-trench $\beta$-Ga$_2$O$_3$ MOSFET with an enlarged view of the trench gate. (b) Multi-trench MOSFET with 8 parallel trenches. (c) Multi-trench MOSFET with 80 parallel trenches.}
    \label{fig3}
\end{center}
\end{figure}

\begin{figure*}
\begin{center}
    \includegraphics[width=\linewidth]{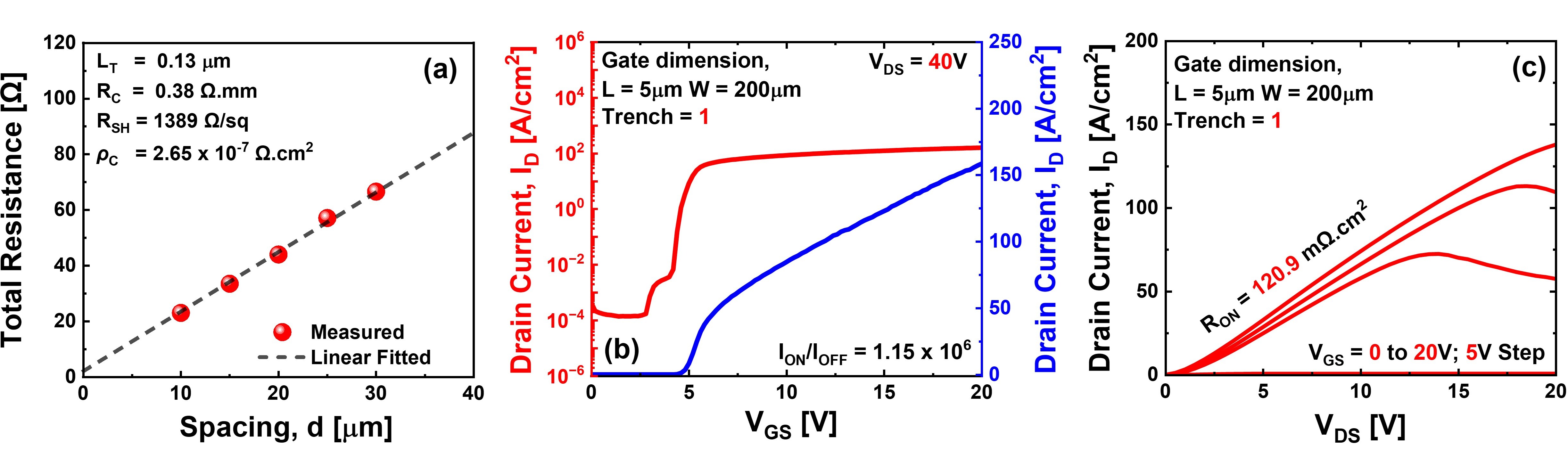}
    \caption{(a) TLM characteristics of the MOCVD-regrown n$^+$ contact layer showing low contact resistivity. (b) Transfer characteristics of the single-trench vertical $\beta$-Ga$_2$O$_3$ MOSFET exhibiting enhancement-mode operation. (c) Output characteristics of the single-trench device.}
    \label{fig4}
\end{center}
\end{figure*}

Unlike conventional approaches employing Si-ion implantation for ohmic contact formation, a heavily Si-doped n$^+$ contact layer was selectively regrown by metal-organic chemical vapor deposition (MOCVD) to form low-resistance source and drain contact regions. The regrowth was performed at 690~$^\circ$C and 60~Torr using triethylgallium (TEGa) and oxygen as the Ga precursor and oxidant source, respectively. The oxygen flow rate was maintained at 400~sccm, while the TEGa molar flow rate was 19.12~$\mu$mol/min. Silane (SiH$_4$) was introduced at a molar flow rate of 145.1~nmol/min, resulting in an approximately 75-nm-thick Si-doped n$^+$ $\beta$-Ga$_2$O$_3$ layer with an electron concentration in the low-$10^{19}$~cm$^{-3}$ range. Fabrication then started with a BCl$_3$/Ar reactive-ion etching (RIE) process on the backside to remove approximately 1~$\mu$m of $\beta$-Ga$_2$O$_3$. A Ti/Au (75/150~nm) ohmic metal stack was subsequently deposited on the backside using electron-beam evaporation, followed by rapid thermal annealing (RTA) in N$_2$ ambient at 470~$^\circ$C for 1~min to form the drain contact.

Top-side Ti/Au/Ni (75/150/15~nm) ohmic source contacts were defined by electron-beam lithography (EBL) and deposited by electron-beam evaporation. Mesa isolation was subsequently performed using a BCl$_3$-based reactive-ion etching (RIE) process. An Al$_2$O$_3$ passivation layer was then deposited by atomic layer deposition (ALD) prior to the gate trench etching process in order to protect the device surface and suppress surface-related leakage current. Subsequently, a U-shaped trench gate structure was formed using RIE. Both the trench and mesa isolation depths were approximately 1000--1100~nm, extending into the unintentionally doped drift layer. Following trench etching, a 50~nm SiO$_2$ gate dielectric layer was conformally deposited by atomic layer deposition (ALD). Finally, Ti/Au/Ni gate metal was deposited using electron-beam evaporation to complete the device fabrication.

Fig.~3(a) shows a top-view SEM image of a single-trench MOSFET together with an enlarged view of the trench gate region. The trench opening width is approximately 5~$\mu$m, and the smooth trench sidewalls indicate good etch quality and dimensional control during the ICP-RIE process. Figures~3(b) and 3(c) show multi-trench MOSFETs consisting of 8 and 80 parallel trench gates, respectively. The multi-trench architecture increases the effective channel width and total current capability while maintaining the same device operating principle. The well-defined trench geometry and uniform spacing observed in the SEM images indicate good lithographic definition and etch control, demonstrating the suitability of the proposed fabrication process for multi-trench vertical $\beta$-Ga$_2$O$_3$ MOSFETs.

\begin{figure*}
\begin{center}
    \includegraphics[width=\linewidth]{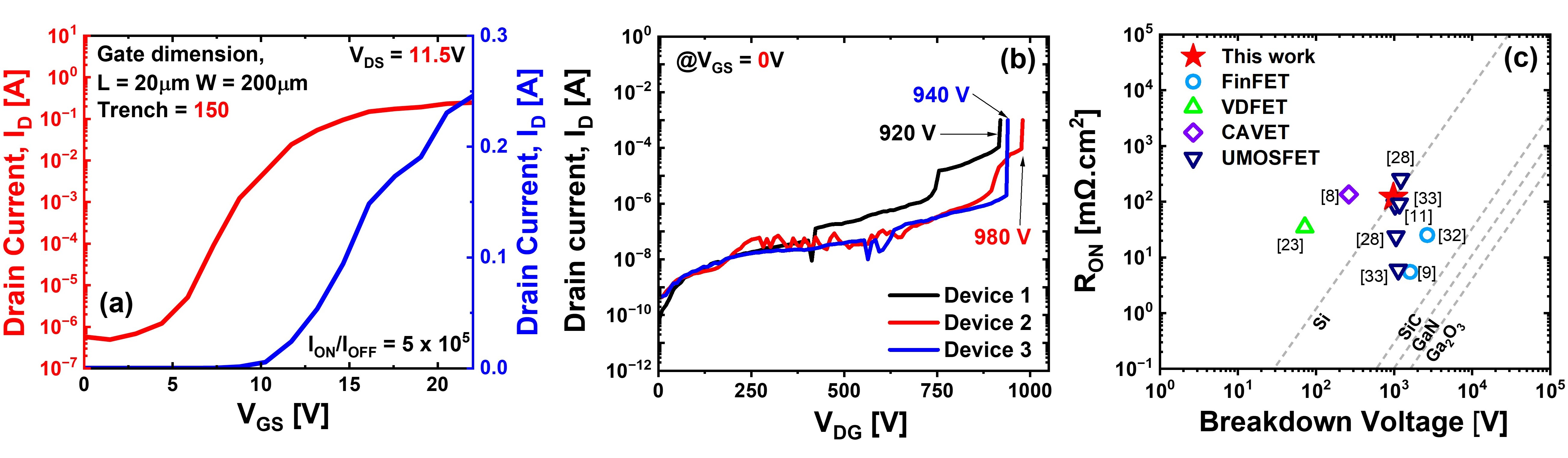}
    \caption{(a) Transfer characteristics of the 150-trench MOSFET demonstrating scalability to larger-area devices. (b) Three-terminal OFF-state breakdown characteristics of the vertical $\beta$-Ga$_2$O$_3$ U-trench MOSFETs measured at $V_{GS}=0$~V. (c) Benchmark of specific ON-resistance ($R_{\mathrm{ON}}$) versus breakdown voltage ($V_{\mathrm{BR}}$) compared with previously reported vertical $\beta$-Ga$_2$O$_3$ FinFETs, VDFETs, CAVETs, and UMOSFETs\cite{ZengMg2022,WongCAVET2019,Xu2025,Liu2024,Hu2019,Li2019,LuoHF2026}.}
    \label{fig5}
\end{center}
\end{figure*}

The ohmic contact properties of the regrown n$^+$ contact layer were evaluated using transmission line model (TLM) measurements, as shown in Fig.~4(a). The measured total resistance exhibited a strong linear dependence on contact spacing, indicating excellent ohmic behavior and uniform current transport across the contact regions. From the linear fit, a transfer length ($L_T$) of 0.13~$\mu$m and a contact resistance ($R_C$) of 0.38~$\Omega\cdot$mm were extracted. The sheet resistance of the regrown n$^+$ layer was determined to be 1389~$\Omega/\square$, corresponding to a specific contact resistivity ($\rho_C$) of $2.65\times10^{-7}~\Omega\cdot$cm$^2$. The low specific contact resistivity demonstrates that selectively MOCVD-regrown n$^+$ contact layers enable excellent ohmic characteristics while eliminating the need for Si-ion implantation and contact activation annealing. These results establish the regrown-contact scheme as a practical alternative for forming low-resistance ohmic contacts in vertical $\beta$-Ga$_2$O$_3$ power MOSFETs \cite{WongNandSi2018,WongCAVET2019,Saha2025}. Although the drift region dominates the ON-resistance of the present high-voltage devices, minimizing contact resistance remains important for reducing parasitic resistance and enabling future device scaling. As future device designs employ smaller cell pitches and reduced channel resistance, the contribution of contact resistance to the total ON-resistance becomes increasingly significant. Therefore, The demonstrated implantation-free regrown-contact process provides a practical alternative to Si-ion implantation for ohmic contact formation in vertical $\beta$-Ga$_2$O$_3$ MOSFETs.

The electrical characteristics of the fabricated enhancement-mode vertical $\beta$-Ga$_2$O$_3$ U-trench MOSFETs are shown in Fig.~4(b)--(c). Fig. 4(b) presents the transfer characteristics measured at $V_{DS}=40$~V. The device exhibited clear normally-OFF behavior with a threshold voltage ($V_{\mathrm{TH}}$) of approximately 5~V, demonstrating the effectiveness of the nitrogen-ion-implanted current blocking layer in suppressing channel conduction at zero gate bias\cite{WongCAVET2019,Ma2023,Liu2024,Zou2025,Xu2025}. The drain current increased rapidly with increasing gate voltage, indicating efficient gate control over channel formation along the trench sidewalls. A peak drain current density of 158 A/cm$^2$ was achieved at high gate bias, highlighting the strong current conduction capability of the vertical device architecture. An ON/OFF current ratio of $1.15\times10^{6}$ was achieved, confirming strong current modulation and low OFF-state leakage current.

Fig. 4(c) shows the output characteristics measured for gate voltages ranging from 0 to 20~V with a step size of 5~V. The device demonstrated well-behaved transistor operation with clear gate modulation and increasing drain current at higher gate bias. The extracted specific ON-resistance ($R_{\mathrm{ON}}$) was 120.9~m$\Omega\cdot$cm$^2$. The relatively higher ON-resistance is primarily attributed to the wide trench opening and current spreading resistance in the drift region. Further reduction of the trench opening width and optimization of the current blocking layer profile are expected to improve channel conductivity and reduce the overall conduction resistance\cite{LiuScaling2026,LuoHF2026}.

To evaluate device scalability, transfer characteristics of a larger-area device containing 150 parallel trench gates are shown in Fig.~5(a). The device exhibited a maximum drain current exceeding 0.25~A while maintaining enhancement-mode operation with an ON/OFF current ratio of approximately $5\times10^5$. The increased total current compared to the single-trench device demonstrates the effectiveness of the multi-trench architecture and confirms that the regrown-contact technology can be extended to larger device geometries without degrading transistor operation. These results highlight the potential of the proposed structure for high-current vertical power switching applications.

Three-terminal OFF-state breakdown characteristics measured at $V_{GS}=0$~V are presented in Fig.~5(b). The breakdown measurements were performed in Fluorinert FC-40 dielectric liquid to suppress air arcing during high-voltage operation. The fabricated devices exhibited breakdown voltages ranging from 920 to 980~V across multiple devices, demonstrating robust blocking capability and good device-to-device reproducibility \cite{Wakimoto2025,Zou2025,Xu2025,Khan2025}. The low OFF-state leakage current prior to breakdown indicates effective carrier depletion by the nitrogen-ion-implanted current blocking layer and good dielectric integrity of the trench-gate structure.

Fig.~5(c) compares the specific ON-resistance and breakdown voltage of this work with previously reported vertical $\beta$-Ga$_2$O$_3$ power transistors, including FinFETs, VDFETs, CAVETs, and UMOSFETs. Although the demonstrated R$_{\mathrm{ON}}$-BV trade-off is comparable to previously reported devices, this work introduces an implantation-free MOCVD-regrown contact technology while achieving one of the lowest reported specific contact resistivities for vertical $\beta$-Ga$_2$O$_3$ MOSFETs. The demonstrated near-1~kV blocking capability together with excellent contact characteristics confirms that low-resistance ohmic contacts can be realized without Si-ion implantation or contact activation annealing, providing a practical fabrication approach for future vertical $\beta$-Ga$_2$O$_3$ power MOSFETs \cite{LiuScaling2026,Yao2026,Mazumder2025}.

In summary, an enhancement-mode vertical $\beta$-Ga$_2$O$_3$ U-trench MOSFET incorporating a nitrogen-ion-implanted current blocking layer (CBL) and MOCVD-regrown n$^+$ contact layers was successfully demonstrated. A multi-energy nitrogen implantation scheme followed by 1100~$^\circ$C annealing enabled the formation of an effective CBL for normally-OFF operation, while the selectively regrown Si-doped n$^+$ contact layers eliminated the need for high-dose Si implantation and activation annealing for ohmic contact formation. The regrown contact structure achieved a low specific contact resistivity of $2.65 \times 10^{-7}~\Omega\cdot$cm$^2$. The fabricated devices exhibited a threshold voltage of approximately 5~V, an ON/OFF current ratio of $1.15\times10^{6}$, a peak current density of 158 A/cm$^2$, and a specific ON-resistance of 120.9 m$\Omega\cdot$cm$^2$. Three-terminal OFF-state breakdown voltages ranging from 920 to 980~V were achieved, demonstrating robust blocking capability and excellent device uniformity. The demonstrated implantation-free regrown-contact process provides a practical, implantation-free contact technology to Si-ion implantation for ohmic contact formation in vertical $\beta$-Ga$_2$O$_3$ MOSFETs. This approach simplifies fabrication by eliminating contact implantation and high-temperature activation annealing while maintaining excellent contact characteristics.

\begin{acknowledgments}
We acknowledge the support from ARPA-E Award No. DE-AR0001879, AFOSR (Air Force Office of Scientific Research) under award FA9550-18-1-0479 (Program Manager: Ali Sayir), from NSF under awards ECCS 223102, 2532899 and Coherent II-VI Foundation Block Gift Program. This work used the electron beam lithography system acquired through NSF MRI award ECCS 1919798.
\dots.
\end{acknowledgments}

\section*{Data Availability Statement}

AIP Publishing believes that all datasets underlying the conclusions of the paper should be available to readers. Authors are encouraged to deposit their datasets in publicly available repositories or present them in the main manuscript. All research articles must include a data availability statement stating where the data can be found. In this section, authors should add the respective statement from the chart below based on the availability of data in their paper.

\nocite{*}
\bibliography{aipsamp}

\end{document}